\documentclass[11pt]{article}

\usepackage{epsf,latexsym,amssymb}
\usepackage{graphicx}
\usepackage{amsmath}  %will give errors with \be \ee !!!

\usepackage{chemarr}
\newcommand{\arrowchem}[1]{\xrightarrow{#1}}
\newcommand{\arrowschem}[2]{\xrightleftharpoons[#1]{#2}}

\newcommand{\comment}[1]{}
\newcommand{\edo}{\end{document}}

\newcommand{\R}{{\mathbb R}}  %ams bold
\newcommand{\Z}{{\mathbb Z}}  %ams bold
\newcommand{\twoif}[4]{
\left\{ \begin{array}{ll}#1&#2\\#3&#4\end{array}\right.
}
\newcommand{\threeif}[6]{
\left\{ \begin{array}{ll}#1&#2\\#3&#4\\#5&#6\end{array}\right.
}
\newtheorem{theorem}{Theorem}
\newtheorem{itlemma}{Lemma}[section] %number by section (set in \em by default)
\newtheorem{itproposition}[itlemma]{Proposition}
\newtheorem{itcorollary}[itlemma]{Corollary}
\newtheorem{itremark}[itlemma]{Remark}
\newtheorem{itdefinition}[itlemma]{Definition}
\newtheorem{itexample}[itlemma]{Example}

\newenvironment{lemma}{\begin{itlemma}\rm}{\end{itlemma}} %no-italics
\newenvironment{remark}{\begin{itremark}\rm}{\end{itremark}} %no-italics
\newenvironment{corollary}{\begin{itcorollary}\rm}{\end{itcorollary}}
\newenvironment{proposition}{\begin{itproposition}\rm}{\end{itproposition}}
\newenvironment{definition}{\begin{itdefinition}\rm}{\end{itdefinition}}
\newenvironment{example}{\begin{itexample}\rm}{\end{itexample}}

\newcommand{\be}[1]{\begin{equation}\label{#1}}
\newcommand{\ee}{\end{equation}}
\newcommand{\bl}[1]{\begin{lemma}\label{#1}}
\newcommand{\ble}[1]{\begin{lemmaex}\label{#1}}
\newcommand{\br}[1]{\begin{remark}\label{#1}}
\newcommand{\bt}[1]{\begin{theorem}\label{#1}}
\newcommand{\bd}[1]{\begin{definition}\label{#1}}
\newcommand{\bp}[1]{\begin{proposition}\label{#1}}
\newcommand{\bc}[1]{\begin{corollary}\label{#1}}
\newcommand{\bex}[1]{\begin{example}\label{#1}}
\newcommand{\ec}{\mybox\end{corollary}}
\newcommand{\eex}{\mybox\end{example}}
\newcommand{\eem}{\mybox\end{example}}
\newcommand{\el}{\mybox\end{lemma}}
\newcommand{\er}{\mybox\end{remark}}
\newcommand{\et}{\qed\end{theorem}}
\newcommand{\ed}{\mybox\end{definition}}
\newcommand{\ep}{\mybox\end{proposition}}
\newcommand{\epr}{\end{proof}}
\newcommand{\bpr}{\begin{proof}}

\newcommand{\ecs}{\end{corollary}}
\newcommand{\eexs}{\end{example}}
\newcommand{\els}{\end{lemma}}
\newcommand{\ers}{\end{remark}}
\newcommand{\ets}{\end{theorem}}
\newcommand{\eds}{\end{definition}}
\newcommand{\eps}{\end{proposition}}
\newcommand{\halmos}{\rule{1ex}{1.4ex}}
\newcommand{\qed}{\hfill \halmos} %put \qed at right margin
\newcommand{\mybox}{\hfill $\Box$} %put \qed at right margin (white square)
\newcommand{\beq}{\begin{eqnarray}}
\newcommand{\eeq}{\end{eqnarray}}
\newcommand{\beqn}{\begin{eqnarray*}}
\newcommand{\eeqn}{\end{eqnarray*}}
\newcommand{\bi}{\begin{itemize}}
\newcommand{\ei}{\end{itemize}}
\newcommand{\ben}{\begin{enumerate}}
\newcommand{\een}{\end{enumerate}}

\newcommand{\bes}[1]{\begin{subequations}\label{#1}\begin{eqnarray}}
\newcommand{\ees}[1]{\end{eqnarray}\end{subequations}}

\newcommand{\st}{\, | \,}

\newenvironment{proof}{\noindent {\em Proof}.\ }{\hspace*{\fill}$\halmos$\medskip}

\newcommand{\mypmatrix}[1]{\left(\begin{array}{cccccccccccc}#1\end{array}\right)}

\newcommand{\rref}[1]{(\ref{#1})}

\title{A technique for determining the signs of sensitivities\\ of steady states
in chemical reaction networks}
\author{Eduardo D. Sontag}

\newcommand{\sx}{\xi }

\newcommand{\xl}{x^{\lambda }}
\newcommand{\sxl}{\sx^{\lambda }}
\newcommand{\pil}{\pi ^{\lambda }}

\newcommand{\pxl}{\frac{\partial \xl}{\partial \lambda }}
\newcommand{\ns}{{n_{\mbox{\tiny \sc s}}}}
\newcommand{\nr}{{n_{\mbox{\tiny \sc r}}}}
\newcommand{\nc}{{n_{\mbox{\tiny \sc c}}}}
\newcommand{\sign}{\mbox{sign}\,}
\newcommand{\signn}{\mbox{sign}}
\newcommand{\vvp}{v^+}
\newcommand{\vvn}{v^-}

\newcommand{\vvpA}{\vvp_A}
\newcommand{\vvnA}{\vvn_A}

\newcommand{\al}{a} %stoichiometry, reactants
\newcommand{\bb}{b} %stoichiometry, products
\newcommand{\VV}{{\mathbf V}}
\newcommand{\VG}{\VV_{G}}
\newcommand{\SV}{{\mathbf S}}
\newcommand{\SG}{\SV_{G}}

\newcommand{\prj}{\frac{\partial R}{\partial x_j}}
\newcommand{\prkj}{\frac{\partial R_k}{\partial x_j}}
\newcommand{\prki}{\frac{\partial R_k}{\partial x_i}}

\newcommand{\Sop}{{\widetilde \Sigma }_0}
\newcommand{\So}{\Sigma _0}
\newcommand{\Sx}{\Sigma }
\newcommand{\signsalpha}{\mu }
\newcommand{\signv}{\widetilde {\mu }}

\begin{document}
\maketitle

\centerline{ABSTRACT}

\medskip

We present a computational procedure to characterize the signs of
sensitivities of steady states to parameter perturbations
in chemical reaction networks.

\section{Introduction}

An important question in the mathematical analysis of chemical reaction
networks is the characterization of sensitivities of steady states to
perturbations in parameters.  An example of a parameter is the total
concentration of an enzyme in its various activity states.  Its value might be
manipulated experimentally in various forms, through expression knock-downs
via interference RNA methods, or up-regulation, titration of inducers,
pharmacological interventions through small-molecule inhibitors, or other
modifications.  Often, one wants to predict the effect of such perturbations,
in a manner that depends only on the structure of the network of reactions and
not on the actual values of other parameters, such as kinetic constants, which
are typically very imperfectly known.

Let us start with a very trivial example.  Suppose that we study the following
reversible bimolecular reaction:
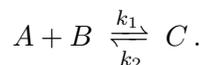
\[
A+B \; \arrowschem{k_2}{k_1} \; C \,.
\]
Let us write lower case letters $a,b,c$ for the concentrations of $A$, $B$,
and $C$ respectively.
Modeling with deterministic mass-action kinetics, the steady states of the
associated ordinary differential equation are obtained by solving
\be{eq:binary_binding}
k_1  a b \,-\,  k_2c \;=\; 0
\ee
subject to two conservation laws:
\[
a+c=A_T \quad \mbox{and} \quad b+c=B_T\,,
\]
where $A_T$ and $B_T$ are two positive constants denoting the total (bound and
unbound) forms of $A$ and $B$ respectively.
For the associated set of ordinary differential equations, all solutions
converge to a unique positive steady state determined
by~\rref{eq:binary_binding} and the conservation laws.

Suppose that we now perform the following experiment.  First, the system is
allowed to relax to steady state, starting from the concentrations $a(0)=A_T$,
$b(0)=B_T$, and $c(0)=0$.  The final concentrations $a_f$, $b_f$, and $c_f$ are
measured.  Next, the experiment is repeated, but the total amount $A_T$ is now
set to a slightly larger value, while $B_T$ is kept constant.  Let us call the
final concentrations obtained in this new experiment, with larger $A_T$, as
$a_f'$, $b_f'$, and $c_f'$.  What can we say about the signs of the differences 
$\Delta a = a_f'-a_f$, $\Delta b = b_f'-b_f$, and $\Delta c = c_f'-c_f$?
One approach to answering this question is to substitute the conservation laws
into the steady state equation~\rref{eq:binary_binding}, for instance
eliminating $a$ and $b$ so that $c=c_f$ can be found by solving the quadratic
equation:
\[
k_1 (A_T-c)(B_T-c) -  k_2c = 0
\]
for the unique root that is between $0$ and $\min\{A_T,B_T\}$:
\[
\frac{k_1(A_T+B_T)+k_2 - \sqrt{(k_1(A_T+B_T)+k_2)^2 - 4 k_1^2 A_TB_T}}{2k_1}
\]
and then $a_f$ and $b_f$ are obtained from $a_f=A_T-c_f$ and $b_f=B_T-c_f$.
A similar solution can be obtained for the larger value of $A_T$, and the
differences $\Delta a$, $\Delta b$, and $\Delta c$ can be computed.  Obviously, this is not a
practical, or even possible, approach for large networks.  On the other hand,
a more conceptual and generalizable approach to this problem is as follows.

Suppose that we view the vector of steady states $x=(a_f,b_f,c_f)$ as a curve
which is parametrized by $A_T$, which we write as an abstract parameter $\lambda $.  
Thus, for all values of this parameter $\lambda $, we have that the following three
equations must hold:
\beqn
k_1  a(\lambda ) b(\lambda ) \,-\,  k_2c(\lambda ) &=& 0\\
a(\lambda )+c(\lambda )&=&\lambda \\
b(\lambda )+c(\lambda )&=&B_T.
\eeqn
Taking derivatives with respect to $\lambda $, we have:
\beqn
k_1  a'(\lambda ) b(\lambda ) + k_1  a(\lambda ) b'(\lambda ) \,-\,  k_2c'(\lambda ) &=& 0\\
a'(\lambda )+c'(\lambda )&=&1\\
b'(\lambda )+c'(\lambda )&=&0.
\eeqn
Substituting $b'(\lambda ) = -c'(\lambda )$ and $a'(\lambda ) =1 -c'(\lambda )$ in the first equation,
we have that:
\[
k_1  (1-c'(\lambda )) b(\lambda ) - k_1  a(\lambda ) c'(\lambda ) \,-\,  k_2c'(\lambda ) \;=\; 0,
\]
which may be re-arranged as:
\[
k_1b(\lambda ) \;=\; M c'(\lambda ),\quad \mbox{where} \;\; M = k_1 a(\lambda ) + k_2 + k_1 b(\lambda ) \,.
\]
Since $M>0$ and $k_1b(\lambda )>0$, we conclude that $c'(\lambda )>0$.  In other words,
$\Delta c>0$ for an increase in $\lambda =A_T$.
Since $b'(\lambda ) = -c'(\lambda )$, we also know that $\Delta b<0$.
What about $\Delta a$?
If we only substitute $b'(\lambda ) = -c'(\lambda )$ in the first equation,
we have that:
\[
k_1  a'(\lambda ) b(\lambda ) = (k_1a(\lambda ) + k_2)c'(\lambda )
\]
and so, using $k_1b(\lambda )>0$ and $k_1a(\lambda ) + k_2>0$, we conclude that $a'(\lambda )$ has
the same sign as $c'(\lambda )$.
Finally, since we also know that $a'(\lambda )+c'(\lambda )=1>0$, this implies that
$a'(\lambda )>0$, so $\Delta a>0$.

The rest of this paper shows how to extend this conceptual argument to
more arbitrary networks.

\section{Preliminaries}

We start with arbitrary systems of ordinary differential equations (ODE's) 
\be{eq:gensys}
\dot x(t) = f(x(t))\,.
\ee
The vectors $x$ are assumed to lie in the positive orthant $\R^\ns_+$ of
$\R^\ns$, that is, $x=(x_1,\ldots ,x_{\ns})^T$ with each $x_i>0$, and $f$ is a
differentiable vector field, mapping $\R^\ns_+$ into $\R^\ns$.
We later specialize to ODE's that describe chemical reaction networks
(CRN's), for which the abstract procedure to be described next can be made
computationally explicit. 
In the latter context, we think of the coordinates $x_i(t)$ of $x$ as
describing the concentrations of various chemical species $S_i$, $i=1,\ldots ,\ns$.

Suppose that $\xl$ describes a $\lambda $-parametrized smooth curve of steady
states for the system~\rref{eq:gensys}, where $\lambda $ is a scalar parameter 
ranging over some open interval $\Lambda $.
The steady state condition amounts to asking that
\be{eq:ss}
f(\xl) = 0
\ee
for all values of the parameter $\lambda \in \Lambda $.

In addition to~(\ref{eq:ss}), we also assume that the steady states of interest
are constrained by a set of algebraic equations
\be{eq:gss}
g_1(\xl) = 0,\; g_2(\xl) = 0,\; \ldots , \ g_{\nc}(\xl) = 0
\ee
where $\nc$ is some positive integer (which we take to be zero when there are
no additional constraints).  We write simply $g(\xl)=0$, where
$g:\R^\ns_+\rightarrow \R^\nc$ is a differentiable mapping whose components are the
$g_i$'s.  Some or all $g_i$ might be linear functions, representing moities or
stochiometric constraints, but nonlinear constraints will be useful when
treating certain examples, as will be discussed later.

Let us denote by 
\[
\sxl \,:=\; \pxl \in  \R^{\ns\times 1}
\]
the derivative of the vector function $\xl$ with respect to $\lambda $, viewed as a
function $\Lambda  \rightarrow \R^{\ns\times 1}$.

We are interested in answering the following question:
\begin{center}
what are the signs of the entries of $\sxl$?
\end{center}
Obviously, the answer to this question will, generally speaking, depend on the
chosen $\lambda $.  
The computation of the steady state $\xl$ as a function of $\lambda $ generally will
involve the numerical approximate solution of nonlinear algebraic equations,
and has to be repeated for each individual parameter $\lambda $.
Our aim is, instead, to provide conditions that allow one to find these signs
independently of the specific $\lambda $, and, even independently of other
parameters that might appear in the specification of $f$ and of $g$, such as
kinetic constants, and to do so using only linear algebraic and logical
operations, with no recourse to numerical approximations.

Proceeding in complete generality, we take the derivative with respect to $\lambda $
in~(\ref{eq:ss}), 
so that, by the chain rule, we have
that $f'(\xl)\sxl = 0$,
where $f'(x)$ denotes the Jacobian matrix of $f$ evaluated at a state $x$.
In other words, 
\be{eq:null_f}
\sxl \in {\cal N}(f'(\xl))\,,
\ee
where ${\cal N}(f'(x))$ denotes the nullspace of the matrix $f'(x)$.
Similarly, we have that
\be{eq:null_g}
\sxl \in {\cal N}(g'(\xl))\,.
\ee
The reason for introducing separately $f$ and $g$ will become apparent later:
we will be asking that each of the $\nc\times \ns$ entries of the Jacobian matrix
of $g$ should not change sign over the state space (which happens, in
particular, when $g$ is linear, as is the case with stoichiometric constraints).
No similar requirement will be made of $f$, but instead, we will study the
special case in which $f$ represents the dynamics of a CRN.

\subsubsection*{Notations for signs of vectors and of subspaces}

For any (row or column) vector $u$ with real entries, we introduce the vector of
signs of entries of $u$, denoted $\sign u$, as the (row or column) vector
with entries in the set $\{-1,0,1\}$ whose $i$th coordinate satisfies:
\[
(\sign u)_i = \threeif%
{-1}{\mbox{if $u_i<0$}}%
 {1}{\mbox{if $u_i>0$}}%
 {0}{\mbox{if $u_i=0$.}}
\]
(The function $\sign$ is sometimes called the ``signature function'' when
viewed as a map $\R^m\rightarrow \{-1,0,1\}^n$.)
More generally, for any subspace ${\cal W}$ of vectors with real entries, we define
\[
\sign {\cal W} = \{\sign v \st v \in  {\cal W}\}\,.
\]
Computing $\sign {\cal W}$ amounts to the combinatorial problem of determining which
orthants are intersected by ${\cal W}$.%
\footnote{We do not need to use this fact, but it is worth noting that, given a
  basis of ${\cal W}$, the signs of ${\cal W}$ represent the ``oriented matroid'' 
  associated to a matrix that lists the basis as its columns, which is the
  set of ``covectors'' of this basis.  This topic is central to the theory of
  oriented matroids.}

We also introduce the positive and negative parts of a vector $u$,
denoted by $u^+$ and $u^-$ respectively, as follows:
\[
(u^+)_i = \twoif
{u_i} {\mbox{if $u_i>0$}}%
{0}   {\mbox{if $u_i\leq 0$}}%
\, \quad\quad
(u^-)_i = \twoif
{-u_i} {\mbox{if $u_i<0$}}%
{0}   {\mbox{if $u_i\geq 0$}\,.}%
\]
Note that $u = u^+ - u^-$,
$\sign u = \sign u^+ - \sign u^-$,  %never used, just FYI
and:
\be{eq:signsign}
(\sign u)^+ = \signn(u^+)\,,\quad
(\sign u)^- = \signn(u^-)\,.
\ee

Suppose that
$u\in \R^{1\times  n}$ and
$v\in \R^{n\times  1}$, for some positive integer $n$.
The equality:
\be{eq:productsign}
\signn(u v) \;=\; \signn\left(\signn(u) \, \signn(v)\right)\,.
\ee
need not hold for arbitrary vectors:
for example, if $u=(1,-1/4,-1/4,-1/4)$ and $v=(1,1,1,1)^T$ 
then $\signn(uv)=\signn(1/4)=1$,
but
\[
\signn\left(\signn(u)\signn(v)\right)=
\signn\left((1,-1,-1,-1)(1,1,1,1)^T\right) = \signn(-2) = -1\,.
\]
However, equality~\rref{eq:productsign} is true provided that
we assume that (a)
$u^-=0$ or $u^+=0$ (that is, either $u_i\geq 0$ for all $i$, or 
$u_i\leq 0$ for all $i$, respectively), and also that (b) $v^-=0$ or $v^+=0$.
This is proved as follows.  
Take first the case $u^-=0$ and $v^-=0$.
Each term in the sum $uv=\sum_{i=1}^nu_iv_i$ is non-negative.
Thus, $uv>0$, that is, $\signn(uv)=1$, if and only if $u_i>0$ and $v_i>0$ for
some common index $i$, 
and $uv = \signn(uv)=0$ otherwise.
Similarly, as $\signn(u) \signn(v)=\sum_{i=1}^n\signn(u_i)\signn(v_i)$,
we know that $\signn(u) \signn(v)>0$, i.e.
$\sign(\signn(u) \signn(v))=1$, if and only if
$\signn(u_i)=\signn(v_i)=1$ for some $i$, and $\signn(u) \signn(v)=0$ otherwise.
But $\signn(u_i)=\signn(v_i)=1$ is the same as $u_i>0$ and $v_i>0$.
Thus~(\ref{eq:productsign}) is true.
The case $u^+=0$ and $v^-=0$ can be reduced to $u^-=0$ and $v^-=0$
by considering $-u$ instead of $u$:
$\signn(u v) = -\signn((-u) v) = -\signn(\signn(-u)\signn(v)) = 
\signn(\signn(u)\signn(v))$.
Similarly for the remaining two cases.

\subsubsection*{A parameter-dependent constraint set}

Denoting
\[
{\cal W}(\xl)={\cal N}(f'(\xl))\bigcap {\cal N}(g'(\xl))
\]
we have that~\rref{eq:null_f} and~\rref{eq:null_g} can be summarized as
follows, in terms of the sign notations just introduced:
\[
\pil\,:= \; \sign \sxl \,\in \, \sign {\cal W}(\xl)\,.
\]
Therefore, one could in principle determine the possible values of $\pil$
once that ${\cal W}(\xl)$ is known.
However, in applications one typically does not know explicitly the curve
$\xl$, which makes the problem difficult because the subspace ${\cal W}(\xl)$ depends
on $\lambda $, and even computing the steady states $\xl$ is a hard problem.
As discussed below, for the special case of ODE systems arising from CRN's, a
more systematic procedure is possible.
Before turning to CRN's, however, we discuss general facts true for all systems.

For every positive concentration vector $x$ define:
\be{eq:sigma}
\Sx(x) \;:= \;
\left\{\signn \left(\nu  f'(x)\right) 
           \st \nu \in \R^{1\times \ns} \right\} 
\bigcup 
\left\{\signn \left(e_i^Tg'(x)\right)
           \st i\in \{1,\ldots ,\nc\} \right\} 
\;\subseteq \; \{-1,0,1\}^{1\times \ns}
\,.
\ee
Here $e_i^T$ denotes the canonical row vector $(0,\ldots 0,,1,0,\ldots 0)$ with a
``$1$'' in the $i$th position and zeroes elsewhere.
The row vectors $\nu $ are used in order to generate an arbitrary linear
combination of the rows of the Jacobian matrix of $f$, a set rich enough to,
ideally, permit the unique determination of the sign of $\sxl$.
As we will use $g$ to introduce constraints of constant sign, and the constant
sign property is not preserved under arbitrary linear combinations of rows, we
only allow $\nu =e_i^T$ for $g$, that is to say, we simply look at the signs of
the rows of $g'(x)$.

Since at a steady state $x=\xl$, $f'(\xl)\sxl=0$ and $g'(\xl)\sxl=0$,
we also have that:
\be{eq:v}
v \,\sxl \;=\; 0
\ee
for each linear combination $v = \nu  f'(\xl)$ and each row $v =  e_i^Tg'(\xl)$.

An easy yet key observation is that the sign vectors in the set $\Sx(\xl)$
strongly constrain the possible signs $\pil=\sign\sxl = \sign \pxl$.
For simplicity in notations, we drop $\lambda $ in $\pil$ and in $\sxl$ when $\lambda $ is
clear from the context, and write simply $\pi $ or $\sx$, with coordinates
$\pi _i$ and $\sx_i$ respectively.

\bl{lem:mainlogical}
Pick any $\lambda \in \Lambda $.
For every $\sigma \in \Sx(\xl)$, and $\pi =\pil$,
it must hold that either:
\be{eq:allzero}
\forall\, i\, \sigma _i\pi _i = 0
\ee
or:
\be{eq:oppsigns}
\left( \exists i\, \sigma _i\pi _i >0\right)
\;\;
\mbox{and}
\;\;
\left( \exists j\, \sigma _j\pi _j <0\right)
\ee
(where $i$ and $j$ range over $\{1,\ldots ,\ns\}$ in all quantifiers).
In other words, either all the coordinates of the vector
\[
\left(\sigma _1\pi _1,\sigma _2\pi _2,\ldots ,\sigma _{\ns}\pi _{\ns}\right)
\]
are zero, or the vector must have both positive and negative entries.
\els

\bpr
Pick $\sigma =\sign v\in \Sx(\xl)$, $\pi =\pil$, $\sx=\sxl$.
Suppose that~(\ref{eq:allzero}) is false.
Then, either there is some $i$ such that $\sigma _i\pi _i >0$
or there is some $j$ such that $\sigma _j\pi _j <0$.
If $\sigma _i\pi _i >0$ for some $i$, then also $v_i\sx_i>0$.
As~(\ref{eq:v}) holds, $\sum_{i=1}^{\ns} v_i\sx_i = 0$, so that there must
exist some other index $j$ for which $v_j\sx_j <0$, which means that
$\sigma _j\pi _j<0$.
Similarly, if there is some $j$ such that $\sigma _j\pi _j <0$, necessarily
there is some $i$ such that $\sigma _i\pi _i >0$, by the same argument.
\epr

We may express the conclusion of Lemma~\ref{lem:mainlogical} in formal logic
terms as follows.  Let $p_{\sigma ,\pi }$ and $q_{\sigma ,\pi }$ be the following logical disjunctions:
\beqn
p_{\sigma ,\pi } &=& \exists i\, \sigma _i\pi _i >0\\
q_{\sigma ,\pi } &=& \exists j\, \sigma _j\pi _j <0
\eeqn
and observe that condition~(\ref{eq:allzero}) is equivalent to asking that
both $p_{\sigma ,\pi }$ and $q_{\sigma ,\pi }$ are false.
Thus, Lemma~\ref{lem:mainlogical} says that, for each $\sigma \in \Sx$, either both 
$p_{\sigma ,\pi }$ and $q_{\sigma ,\pi }$ are false or both $p_{\sigma ,\pi }$ and $q_{\sigma ,\pi }$ are
true.
The ``XNOR($p$,$q$)'' binary function has value ``true'' if and only if $p$ and
$q$ are simultaneously true or false. 
Thus, Lemma~\ref{lem:mainlogical} asserts that this logical statement is true,
for $\pi =\pil$:
\be{eq:xnor}
\mbox{XNOR}(p_{\sigma ,\pi },q_{\sigma ,\pi })
\quad
\forall\, \sigma \in \Sx\,.
\ee
Given any two sign vectors $\sigma $, $\pi $, testing this property is simple in any
programming language.  For example, in MATLAB{\textregistered} syntax, one may
write:
\beqn
\zeta  &=& \sigma .*\pi \\
 p &=& \mbox{sign\,}(\mbox{sum\,} (\zeta >0))\\
 q &=& \mbox{sign\,}(\mbox{sum\,} (\zeta <0))\\
\mbox{XNOR} &=& \mbox{sign\,} (p*q + (1-p)*(1-q))
\eeqn
and the variable XNOR will have value $1$ if $\mbox{XNOR}(p_{\sigma ,\pi },q_{\sigma ,\pi })$
is true, and value $0$ otherwise.

The basis of our approach will be as follows.  We will show how to obtain a
state-independent set $\So$ which is a subset of $\Sx(x)$ for all states $x$.
In particular, for all steady states $\xl$, we will have:
\be{eq:sigma0_in_all_sigma}
\So \; \subseteq  \; \bigcap_{\lambda \in \Lambda } \Sx(\xl) \,.
\ee
Compared to the individual sets $\Sx(\xl)$, which depend on the particular steady
state $\xl$, the elements of this subset are obtained using only linear 
algebraic operations; the computation of $\So$ does not entail solving
nonlinear equations nor simulating differential equations.
Once that this set $\So$ (or even just some large subset of it, which is
easier to compute) has been obtained, we may ask, for each potential
sign vector $\pi $, if~\rref{eq:xnor} is true or not.
Thus, for each $\pi $, we need to test if the conjunction of the clauses
in~\rref{eq:xnor}: 
\be{eq:conjunction}
\bigwedge_{\sigma \in \So} \mbox{XNOR}(p_{\sigma ,\pi },q_{\sigma ,\pi })
\ee
(or the conjunction only over a more easily computed subset) is true or false.
In other words, we are interested in computing the subset of sign vectors $\pi $
for which~\rref{eq:conjunction} is valid.  This question is one of propositional
logic (there are only $3^{\ns}$ possible sign vectors), and as such is
decidable algorithmically, although it has large computational complexity.

We prefer to carry out a sieve procedure for restricting the possible sign
vectors, by testing each $\pi $ one at a time.  For moderate numbers of species,
this is easy and fast to perform computationally.  So we test for each $\pi $
if~\rref{eq:conjunction} is valid.  
If false, then the sign vector $\pi $ is ruled out as a possible sign and
eliminated from the list.
The surviving $\pi $'s are the possible sign vectors.
Of course, since~\rref{eq:xnor} is only a necessary, and not a sufficient,
condition, we are not guaranteed to find a minimal set of signs.
However, we find for many examples that the procedure indeed leads to a
unique, or close to unique, solution, after deleting the zero solution
(since $\sigma =0$ is always a solution)
and also deleting one element in the pair $\{\sigma ,-\sigma \}$ for
each $\sigma $ (since $\nu \sx=0$ implies $\nu (-\sx)=0$, solutions appear always in
pairs).

Testing~\rref{eq:conjunction}, for a fixed $\pi $, is itself a hard
computational problem (NP-hard on the number of species) and hence infeasible
for large-scale networks.  Good heuristics, such as the
Davis-Putnam-Logemann-Loveland (DPLL) algorithm for clauses in conjunctive
normal form, are extensively discussed in the rich literature on satisfiability.
%end comments
However, we have found that a straightforward exhaustive testing of all
possibilities is quite useful, as long as the number of species is reasonably
small. 

The key issue, then, is to find a way to explicitly generate a state-independent
subset $\So$ of $\Sx(\xl)$, and we turn to that problem next.

\section{CRN terminology and notations}

We consider a collection of chemical reactions that involves a
set of $\ns$ ``species'':
\[
S_i , \; i \in \{1,2, \ldots \ns \} \,.
\]
The ``species'' might be ions, atoms, or large molecules, depending on the
context.
A \emph{chemical reaction network} (``CRN'' for short) involving these species
is a set of chemical reactions $\mathcal{R}_j$, $j\in \{1,2, \ldots , \nr \}$,
represented symbolically as:
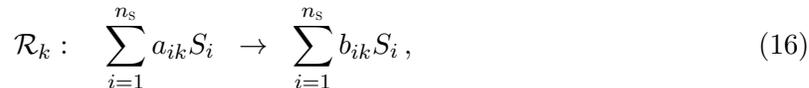
\be{reactions}
 \mathcal{R}_k: \quad \sum_{i =1}^{\ns} \al_{ik} S_i \;\;\rightarrow \;\;
                      \sum_{i =1}^{\ns} \bb_{ik} S_i \,,
\ee
where the $\al_{ik}$ and $\bb_{ik}$ are some non-negative integers that quantify
the number of units of species $S_i$ consumed, respectively produced, by
reaction ${\cal R}_k$.
Thus, in reaction 1, $\al_{11}$ units of species $S_1$ combine with
$\al_{21}$ units of species $S_2$, etc., to produce $\bb_{11}$ units of
species $S_1$, $\bb_{21}$ units of species $S_2$, etc., and similarly for
each of the other $\nr-1$ reactions.

We will assume the following ``non autocatalysis'' condition: no
species $S_i$ can appear on both sides of the same reaction.
With this assumption, either $\al_{ik}=0$ or $\bb_{ik}=0$ for each 
species $S_i$ and each reaction ${\cal R}_k$ (both are zero if the species in
question is neither consumed nor produced),
Note that we are not excluding autocatalysis which occurs through one ore more
intermediate steps, such as the autocatalysis of $S_1$ in  
$S_1+S_2\rightarrow S_3\rightarrow 2S_1+S_4$, so this assumption is not as
restrictive as it might at first appear. 

Suppose that $\al_{ik}>0$ for some $(i,k)$; then we say that species $S_i$ is a
\emph{reactant} of reaction ${\cal R}_k$, and by the non autocatalysis assumption,
$\bb_{ik}=0$ for this pair $(i,k)$.
If instead $\bb_{ik}>0$, then we say that species $S_i$ is a \emph{product} of
reaction ${\cal R}_k$, and again by the non autocatalysis assumption, $\al_{ik}=0$ for
this pair $(i,k)$.

It is convenient to arrange the $\al_{ik}$'s and $\bb_{ik}$'s into two
$\ns\times \nr$ matrices $A$, $B$ respectively, and introduce
the \emph{stoichiometry matrix} $\Gamma  = B-A$.  In other words,
\[
\Gamma  = \left(\gamma _{ij}\right)_{ij} \in \R^{\ns\times \nr}
\]
is defined by:
\begin{equation}
\label{stocmatrix}
\gamma _{ij} \;=\; \bb_{ij}-\al_{ij}\,,\quad
i=1,\ldots ,\ns\,, \quad
j=1,\ldots ,\nr\,.
\end{equation}
The matrix $\Gamma $ has as many columns as there are reactions.
Its $k$th column shows, for each species (ordered according to their index $i$),
the net ``produced$-$consumed'' by reaction ${\cal R}_k$.
The symbolic information given by the reactions~(\ref{reactions}) is
summarized by the matrix $\Gamma $.
Observe that $\gamma _{ik}=-\al_{ik}<0$ if $S_i$ is a reactant of reaction ${\cal R}_k$,
and $\gamma _{ik}=\bb_{ik}>0$ if $S_i$ is a product of reaction ${\cal R}_k$.

To describe how the state of the network evolves over time, one must provide
in addition to $\Gamma $ a rule for the evolution of the vector:
\[
\mypmatrix{[S_1(t)] \cr [S_2(t)] \cr \vdots \cr [S_{\ns}(t)]}\,,
\]
where the notation $[S_i(t)]$ means the concentration of the species $S_i$ at
time $t$.
We will denote the concentration of $S_i$ simply as
$x_i(t) = [S_i(t)]$ and let $x=(x_1,\ldots ,x_{\ns})^T$.
Observe that only non-negative concentrations make physical sense.
A zero concentration means that a species is not present at all; we will be
interested in \emph{positive vectors} $x$ of concentrations, those for which
$x_i>0$ for all $i$, meaning that all species are present.

Another ingredient that we require is a formula for the actual rate at which
the individual reactions take place.
We denote by $R_k(x)$ be algebraic form of the $k$th reaction.
We postulate the following two axioms that the reaction rates $R_k(x)$,
$k=1,\ldots ,\nr$ must satisfy:
\bi
\item
for each $(i,k)$ such that species $S_i$ is a reactant of ${\cal R}_k$,
$\prki(x)>0$ for all (positive) concentration vectors $x$;
\item
for each $(i,k)$ such that species $S_i$ is not a reactant of ${\cal R}_k$,
$\prki(x)=0$ for all (positive) concentration vectors $x$.
\ei
These axioms are natural, and are satisfied by every reasonable model,
and specifically by mass-action kinetics, in which the reaction rate is
proportional to the product of the concentrations of all the reactants:
\[
R_k(x) = \kappa _k \prod_{i=1}^{\ns} x_i^{\al_{ij} }
\mbox{ for all } j=1,\ldots ,\nr
\]
(the positive coefficients $\kappa _k$ are the reaction, or kinetic, constants;
$x_i^{\al_{ij}}=1$ when $\al_{ij}=0$).

Recall that $\al_{ik}>0$ and $\bb_{ik}=0$ if and only if 
$S_i$ is a reactant of ${\cal R}_k$.
Therefore the above axioms state that, for every positive $x$,
\be{RiffG}
\prki(x)>0 \;\Longleftrightarrow\; \al_{ik}>0
\ee
and also
\be{RiffG0}
\prki(x)=0 \;\Longleftrightarrow\; \al_{ik}=0
\ee
because the expressions on both sides are either zero or positive.

We arrange reactions into a column vector function $R(x)\in \R^{\nr}$:
\[
R(x):= 
\mypmatrix{R_1(x) \cr R_2(x) \cr \vdots \cr R_{\nr}(x)} \,.
\]
With these conventions, the system of differential equations associated to the
CRN is given as follows: 
\be{chemreactionnetwork}
\frac{dS}{dt} \;=\; f(x) \;=\; \Gamma \, R(x) \,.
\ee
Observe that $f'(x) = \Gamma  R'(x)$, where $R'(x)$ is the Jacobian matrix of $R$,
which is the matrix whose $(k,j)$th entry is $\prkj(x)$.

We will assume from now also specified a differentiable mapping
\[
g \,:\; \R^\ns_+ \rightarrow  \R^\nc \,,
\]
where $\nc$ is some positive integer (possibly zero,
to indicate the case where there are no additional constraints), and $g$ has
the property that 
\be{eq:assume_constant_signs_g}
\mbox{all $\nc\times \ns$ entries of the Jacobian matrix $g'(x)$ have
constant sign.}
\ee
This happens in the special case when $g$ is linear, as is the case for
stoichiometric constraints.  
It is perfectly fine to add linear combinations of those rows of $g$ that are
linear, since that will not change the constant sign assumption on $g'$.
We assume in the theoretical discussion that $g$ has been extended by possibly
adding one or more such combinations.
Observe that a nonlinear $g$ may also have the constant sign property.  For
example,
suppose that $\ns=5$, $\nc=1$,
and 
\[
g(x) = a x_1x_3- b x_2^2
\]
where $a$ and $b$ are positive constants.
Then the Jacobian matrix (gradient, since $\nc=1$) is:
\[
g'(x) = \nabla g(x) = (a x_3 \,,\, -2b x_2 \,,\, a x_1 \,,\, 0 \,,\, 0)
\]
which has constant sign $(1,-1,1,0,0)$.

For chemical reaction networks, it is not necessary for the entries of $f'(x)$,
and much less the entries of the products $\nu f'(x)$ for vectors $\nu $, to have
constant sign.
Our next task will be to introduce algebraic conditions that allow one to
check if the sign is constant, for any given vector $\nu $.
Before proceeding, however, we give an example of non-constant sign.
Take the following CRN, with $\ns=4$ and $\nr=2$:
\be{counterexample:R1R2}
 \mathcal{R}_1: \; X_1+X_2 \rightarrow  X_4\,,\quad\quad
 \mathcal{R}_2: \; X_2+X_3 \rightarrow  X_1
\ee
which is formally specified, assuming mass-action kinetics, as follows:
\[
A = \mypmatrix{%
1 & 0 \cr
1 & 1 \cr
0 & 1 \cr
0 & 0}\,,\quad
B = \mypmatrix{%
0 & 1 \cr
0 & 0 \cr
0 & 0 \cr
1 & 0}\,,\quad
\Gamma  = \mypmatrix{%
-1 & 1 \cr
-1 & -1 \cr
 0 & -1 \cr
 1 & 0}\,,\quad
R(x) = (k_1x_1x_2,k_2x_2x_3)^T \,.
\]
Thus the ODE set $\dot x=f(x)=\Gamma R(x)$ corresponding to this CRN has:
\beqn
f(x) \;=\; \mypmatrix{%
 -k_1x_1x_2 + k_2x_2x_3 \cr
 -k_1x_1x_2 - k_2x_2x_3 \cr
- k_2x_2x_3 \cr
  k_1x_1x_2} \,.
\eeqn
Let $\nu =e_1^T$.
Observe that $\nu f'(x) = (-k_1x_2,-k_1x_1+k_2x_3,k_2x_2,0)$ does not have
constant sign, because its second entry, which is the same as the $(1,2)$
entry of $f'(x)$, is the function $-k_1x_1+k_2x_3$, which changes sign
depending on whether $x_1>k_2x_3/k_1$ or $x_1<k_2x_3/k_1$.
Ruling out vectors $\nu $ that lead to such ambiguous signs is the purpose of
our algorithm to be described next.

\section{Sensitivities for CRN's}

Introduce the following space:
\[
\VV \;:=\;\mbox{row span of }\Gamma \;=\;\left\{\nu \Gamma  \st 
                            \nu  \in  \R^{1\times \ns} \right\} \; \subseteq \; \R^{1\times \nr}
\,.
\]
Since $f'(x) = \Gamma  R'(x)$, the definition~\rref{eq:sigma} of $\Sx$ becomes:
\[
\Sx(x) \;:=\; \left\{\signn\left(vR'(x)\right) \st v\in \VV \right\}
\bigcup 
\left\{\sign \left(e_i^Tg'(x)\right)
           \st i\in \{1,\ldots ,\nc\} \right\} 
\;\subseteq \; \{-1,0,1\}^{1\times \ns}
\]
when specialized to CRN.

As we assumed Property~\rref{eq:assume_constant_signs_g}, the expressions
$\sign (e_i^Tg'(x))$ are actually independent of $x$.
On the other hand, the sign vectors $\sigma =\sign v R'(x)$ generally depend on the
particular $x$.  
The following Lemma shows that, for vectors $\rho $ with non-negative entries,
the sign of the vector $\rho  R'(x)$ is the same, no matter what the state $x$ is,
and moreover, this sign can be explicitly computed using only stoichiometry
information. 
We denote by 
\[
A_j=(\al_{j1},\ldots ,\al_{j\nr})^T \;\in \; \R^{\nr\times 1}
\]
the $j$th column of the transpose $A^T$, i.e.. the
transpose of the $j$th row of $A$.

\bl{lem:rhoA}
For any positive concentration vector $x$, any non-negative row vector $\rho $
of size $\nr$, and any species index $j\in \{1,\ldots ,\ns\}$:
\be{rhoA0}
\rho  A_j = 0  \;\Longleftrightarrow\; \rho  \prj(x) = 0\,.
\ee
Thus, also
\be{rhoA}
\rho  A_j > 0  \;\Longleftrightarrow\; \rho  \prj(x) > 0 \,,
\ee
since the expressions in each side of~\rref{rhoA0} can only be zero or positive.
\els
\bpr
We have that
\[
\rho  A_j = \sum_{k\in K_\rho } \rho _k \al_{jk}
\]
where $K_\rho := \{k | \rho _k>0\}$.
Since every $\al_{jk}\geq 0$, the equality $\rho  A_j=0$ holds if and only if
$\al_{jk}=0$ for all $k\in K_\rho $.
Similarly, from
\[
\rho  \prj(x) = \sum_{k\in K_\rho } \rho _k \prkj(x)
\]
and $\prkj(x)\geq 0$
we have that $\rho  \prj(x) = 0$ if and only if $\prkj(x)=0$ for all $k\in K_\rho $.
From~(\ref{RiffG0}), we conclude~(\ref{rhoA0}).
\epr

Lemma~\ref{lem:rhoA} is valid for all non-negative $\rho $.  When specialized to
$v=\nu \Gamma \in \VV$, and defining $\sigma  = \sign vR'(x)$, it says that $\sigma $ does not
depend on $x$.  However, elements of the form $v=\nu \Gamma \in \VV$ will generally not
be non-negative (nor non-positive), so the lemma cannot be applied to them.
Instead, we will apply Lemma ~\ref{lem:rhoA} to the positive and negative
parts of such a vector, but only when such positive and negative parts satisfy
a certain ``orthogonality'' property, as defined by the subset of $\VV$
introduced below. 

\subsubsection*{A state-independent subset of $\Sx$}

For any $v\in \VV$, consider the sign vector
$\signv_v := \sign v A^T \in  \{-1,0,1\}^{1\times \ns}$,
whose $j$th entry is $vA_j = \nu \Gamma A_j$ if $v=\nu \Gamma $ with $\nu \in \R^{1\times \ns}$,
as well as the positive and negative parts of $v$, $\vvp$ and $\vvn$,
Define the following set of vectors (``$G$'' for ``good''):
\[
\VG\;:=\; \left\{
v\in \VV
\,|\, \mbox{for each }j\in \{1,\ldots ,\ns\}
\mbox{ either }
             \vvp A_j = 0
\mbox{ or }
             \vvn A_j = 0
\right\}\,.
\]
Observe that, if 
$v\in \VG$,
then
\be{cases_signs}
vA_j = (v^+ - v^-)A_j = v^+A_j - v^-A_j =
\threeif  {v^+A_j}{\mbox{if } v^-A_j = 0}%
        {-v^-A_j}{\mbox{if } v^+A_j = 0}%
        {0}{\mbox{if } v^+A_j = v^-A_j =0 \,.}%
\ee
Consider the following set of sign vectors $\signv_v$ parametrized by elements of $\VG$:
\be{eq:sigma0}
\Sop \; := \; \left\{\signv_v = 
\signn(vA^T) \st v
\in  \VG\right\} 
      \; \subseteq \; \{-1,0,1\}^{1\times \ns}\,.
\ee
The key fact is that this is a subset of $\Sx(x)$
for all $x$, as shown next.

\bl{lem:uniquesign}
For every positive concentration vector $x$,
\[
\Sop \subseteq  \Sx(x).
\]
\els
\bpr
Pick any 
$\signv_v\in \Sop$, where $v\in \VG\subseteq \VV$,
and fix any positive concentration vector $x$.
We must prove that 
$\signv_{v}\in \Sx(x)$.
As $\Sx(x)$ includes all expressions of the form
$\signn(vR'(x))$, for $v\in \VV$,
it will suffice to show that, for this same vector $v$,
\be{eq:uniquesign}
\signn\left(v\prj(x)\right) = \signn \left(vA_j\right)
\ee
for each species index $j\in \{1,\ldots ,\ns\}$.
For each $j\in \{1,\ldots ,\nr\}$, we will show the following three statements:
\be{eq:uniquesign1}
\vvn A_j>0 \mbox{ (and so } \vvp A_j = 0\mbox{)} \;\;\Longrightarrow\;\;
v\prj(x) = -\vvn\prj(x) < 0 \,,
\ee
\be{eq:uniquesign2}
\vvp A_j>0 \mbox{ (and so } \vvn A_j = 0\mbox{)}  \;\;\Longrightarrow\;\;
v\prj(x) = \vvp\prj(x) > 0 \,,
\ee
and
\be{eq:uniquesign3}
\vvn A_j = \vvp A_j = 0 \;\;\Longrightarrow\;\;
v\prj(x) = 0 \,.
\ee
Suppose first that $\vvn A_j>0$.
Applying ~(\ref{rhoA0})
with $\rho =\vvp$, we have that $\vvp\prj(x) = 0$.
Applying ~(\ref{rhoA})
with $\rho =\vvn$, we have that $\vvn\prj(x) > 0$.
Therefore 
\[
v\prj(x) 
= (\vvp - \vvn)\prj(x) = \vvp\prj(x) - \vvn\prj(x) = - \vvn\prj(x) < 0\,,
\]
thus proving~(\ref{eq:uniquesign1}).
If, instead, $\vvn A_j =0$ and $\vvp A_j > 0$, a similar argument shows
that~(\ref{eq:uniquesign2}) holds.
Finally, suppose that $\vvp A_j = \vvn A_j =0$.
Then, again by~(\ref{rhoA0}), applied to $\rho =v^+$ and $\rho =v^-$,
\[
v\prj(x) 
= (\vvp - \vvn)\prj(x) =  0\,,
\]
and so~(\ref{eq:uniquesign3}) holds.
The desired equality~\rref{eq:uniquesign} follows
from~\rref{eq:uniquesign1}-\rref{eq:uniquesign3}.
Indeed, we consider three cases:
(a) $v A_j < 0$,
(b) $v A_j > 0$, and
(c) $v A_j = 0$.
In case (a), \rref{cases_signs} shows that 
$v A_j = -v^-A_j$
(because the first and third cases would give a non-negative value),
and therefore  $-v^-A_j<0$, that is, $v^-A_j>0$,
so~\rref{eq:uniquesign1} gives that 
$v\prj(x)$
is also negative.
In case (b), similarly $\vvp A_j=v A_j>0$, 
and so~\rref{eq:uniquesign2}
shows~\rref{eq:uniquesign}. 
Finally, consider case (c), 
$v A_j = 0$.
If it were the case that $\vvp A_j$ is nonzero, then, since $v\in \VG$,
$\vvn A_j=0$, and therefore~\rref{cases_signs} gives that
$v A_j=v^+A_j>0$, 
a contradiction; similarly, $\vvn A_j$ must also be zero.
So,~\rref{eq:uniquesign3} gives that 
$v\prj(x)=0$
as well.
\epr

\br{rem:interpretM}
To interpret the set $\VG$, it is helpful to study the special case in which
$v$ is simply a row of $\Gamma $, that is, $v=\nu \Gamma $ and
$\nu =e_i^T$, the canonical row vector $(0,\ldots 0,,1,0,\ldots 0)$ with a ``$1$''
in the $i$th position and zeroes elsewhere.
Since
\[
e_i^TB-e_i^TA = e_i^T(B-A) = e_i^T\Gamma  = \vvp - \vvn \,,
\]
and the vectors $e_i^TB$ and $e_i^TA$ have non-overlapping positive entries
(by the non autocatalysis assumption),
we have that $\vvp=e_i^TB$ and $\vvn=e_i^TA$.
Since $e_i^TBA_j =\sum_k \bb_{ik}\al_{jk}$, asking that this number be
positive amounts to asking that
\be{eq:ij1}
\mbox{$i$ is a product of some reaction ${\cal R}_{k}$ which has $j$ as a reactant.}
\ee
Since $e_i^TAA_j =\sum_k \al_{ik}\al_{jk}$, asking that this number is
positive amounts to asking that
\be{eq:ij2}
\mbox{$i$ and $j$ are both reactants in some reaction ${\cal R}_{k'}$.}
\ee
Thus, if the network in question has the property that~(\ref{eq:ij1})
and~(\ref{eq:ij2}) cannot both hold simultaneously for any pair of species
$i,j$, then we cannot have that both $e_i^TBA_j >0$ and $e_i^TAA_j >0$ hold.
In other words, $e_i^T\in \VG$ for all $i$.

As an illustration, take the CRN
$\mathcal{R}_1: X_1+X_2 \rightarrow  X_4$ and $\mathcal{R}_2: X_2+X_3 \rightarrow  X_1$
treated in~\rref{counterexample:R1R2}.
We claim that $e_1^T\not\in \VG$, which reflects the fact that $e_1^Tf'(x)$
does not have constant sign.
Indeed, in this case we have that, with $i=1$ and $j=2$, $X_1$ and $X_2$ are
reactants in $\mathcal{R}_1$ but 
$X_1$ is also a product of reaction ${\cal R}_{2}$, which has $X_2$ as a reactant.
Algebraically, $e_1^T\Gamma  = (-1,1) = (0,1)-(1,0) = \vvp - \vvn$ and $A_2=(1,1)^T$,
so $\vvp A_2=1$ and $\vvn A_2 = 1$.
This means that $\nu =e_1^T\not\in \VG$, since the property defining $\VG$ would
require that at least one of $\vvp A_2$ or $\vvn A_2$ should vanish.
We have re-derived, in a purely algebraic manner, the fact that 
$-k_1x_1+k_2x_3$ changes sign.
\er

Testing whether a given vector 
$v\in \VV$, $v=\nu \Gamma $ with $\nu \in  \R^{1\times \ns}$,
belongs to $\VG$
is easy to do.  For example, in MATLAB\textregistered-like syntax, one may
write:
\beqn
   v &=& \nu  * \Gamma \\
\vvp &=& (v>0).*v\\
\vvn &=& -(v<0).*v\\
\vvpA &=& \signn(\vvp * A')\\
\vvnA &=& \signn(\vvn * A')
\eeqn
and we need to verify that the vectors $\vvpA$ and $\vvnA$ have disjoint
supports, which can be done with the command
\[
\mbox{sum}(\vvpA.*\vvnA)=%
=0
\]
which returns $1$ (true) if and only if 
$v\in \VG$,
in which case we
accept $v$ and we may use
$\sigma  = \signn\left(vA^T\right)$
to test the conditions in
Lemma~\ref{lem:mainlogical}.

\subsubsection*{Explicit generation of elements of $\Sop$}

The set $\Sop$ defined in~(\ref{eq:sigma0}) is constructed in such a way
as to be independent of states $x$, which makes it more useful than the sets
$\Sx(x)$ from a computational standpoint.  Yet, in principle, computing this
set potentially involves the testing of the conditions ``$\vvp A_j=0$ or 
$\vvn A_j=0$'' that define the set $\VG$, for every $v=\nu \Gamma $, that is,
for every possible real-valued vector $\nu \in \R^{1\times \ns}$ (and each $j$).
We describe next a more combinatorial way to generate the elements of $\Sop$.

We introduce the set of signs associated to the row span $\VV$ of $\Gamma $:
\be{def:signV}
\SV\;:=\; \sign \VV \; \subseteq \; \{-1,0,1\}^{1\times \ns} \,.
\ee
Denote:
\[
\alpha \;:=\; \sign A^T
\in  \{0,1\}^{\nr\times \ns}
\]
so that the $j$th column of $\alpha $ is $\alpha _j =\sign A_j\in \{0,1\}^{\nr\times 1}$.

\bl{lem:signscombinatorial}
Pick any $s\in \SV$,
$s=\sign v$, where 
$v\in \VV$.
Then, for each $j\in \{1,\ldots ,\ns\}$:
\[
\signn(\vvp A_j) \;=\; \signn(s^+\alpha _j)\,, \quad \quad
\signn(\vvn A_j) \;=\; \signn(s^-\alpha _j) \,.
\]
\els

\bpr
By (\ref{eq:productsign}), applied with $u=\vvp$ and $v=A_j$,
$\signn(\vvp A_j)= \signn(\signn(\vvp)\alpha _j)$.
By (\ref{eq:productsign}) applied with $u=\vvn$ and $v=A_j$,
$\signn(\vvn A_j)=\signn(\signn(\vvn)\alpha _j)$.
Since, by~(\ref{eq:signsign}) applied with $u=v$,
$s^+ = \signn(v^+)$ and $s^- = \signn(v^-)$, the conclusion follows.
\epr

In analogy to the definition of the set $\VG$, we define (``$G$'' for ``good''):
\[
\SG\;:=\; \left\{s \in  
\SV
\,|\, \mbox{for each }j\in \{1,\ldots ,\ns\}
\mbox{ either }
             s^+ \alpha _j = 0
\mbox{ or }
             s^- \alpha _j = 0
\right\}\,.
\]
Observe that, if $s\in \SG$, then
\be{cases_signs_0}
s \alpha _j = (s^+ - s^-)\alpha _j = s^+a_j - s^-a_j =
\threeif  {s^+\alpha _j}{\mbox{if } s^-\alpha _j = 0}%
        {-s^-\alpha _j}{\mbox{if } s^+\alpha _j = 0}%
        {0}{\mbox{if } s^+\alpha _j = s^-\alpha _j =0 \,.}%
\ee

Consider the following set of sign vectors parametrized by elements of
$\SG$:

\be{eq:sigma01}
\So \; := \; \left\{\signsalpha_s = \signn(s \alpha ) \st s \in  \SG\right\} 
      \; \subseteq \; \{-1,0,1\}^{1\times \ns}\,.
\ee

\bp{cor:combinatorialSo_elements}
Pick any $s\in \SV$, $s=\sign v$, where $v\in \VV$.
Then
\[
s\in \SG \;\;\mbox{if and only if} \;\; v\in \VG
\]
and for such $s$ and 
$v$,
\be{equality_signs}
\signn(vA^T) \;=\; \signn(s\alpha ) \,.
\ee
\eps

\bpr
Let $s=\sign v$, $v\in \VV$,
and pick any $j\in \{1,\ldots ,\ns\}$.
We claim that $s^{\pm}\alpha _j=0$ if and only if $v^{\pm}A_j=0$.
Since $j$ is arbitrary, this shows that $s\in \SG$ if and only if $v\in \VG$.
Indeed, suppose that $s^+\alpha _j=0$.
By Lemma~\ref{lem:signscombinatorial},
$\signn(\vvp A_j)=\signn(s^+\alpha _j)=0$,
so $\vvp A_j=0$.
Conversely, if $\vvp A_j=0$ then $s^+\alpha _j=0$, for the same reason.
Similarly, $s^-\alpha _j=0$ is equivalent to $\vvn A_j=0$.

Suppose now that $s\in \SG$ and $v\in \VG$, and pick any $j\in \{1,\ldots ,\ns\}$.
Assume that $s^+\alpha _j=0$.
Since, by~\rref{cases_signs_0} and~\rref{cases_signs}, $s\alpha _j=-s^-\alpha _j$ and
$vA_j=-v^-A_j$,
we have, again by Lemma~\ref{lem:signscombinatorial}, that
\[
\signn(s\alpha _j) = -\signn(s^-\alpha _j) = -\signn(\vvn A_j) = \signn(vA_j)\,.
\]
If, instead, $s^-\alpha _j=0$ (and thus $\vvn A_j=0$),
\[
\signn(s\alpha _j) = \signn(s^+\alpha _j) = \signn(\vvp A_j) = \signn(vA_j)\,.
\]
As $j$ was arbitrary, and we proved that the $j$th coordinates of the two
vectors in~\rref{equality_signs} are the same, the vectors must be the same.
\epr

\bc{lem:combinatorialSo}
$\Sop=\So$.
\ecs

\bpr
Pick any element of $\Sop$, $\signv_v = \signn(vA^T)$, $v\in \VG$.
By Corollary~\ref{cor:combinatorialSo_elements},
$s=\sign v \in  \SG$.
Moreover, also by Corollary~\ref{cor:combinatorialSo_elements},
$\signv_v = \signn(s\alpha )$, so we know that $\signv_v\in \So$.
Conversely, take an element $\signsalpha_s\in \So$.
This means that $\signsalpha_s = \signn(s \alpha )$ for some $s \in  \SG\subseteq \SV=\sign\VV$.
Let $v\in \VV$ be such that $s=\sign v$.
By Corollary~\ref{cor:combinatorialSo_elements}, $v\in \VG$, and also
$\signsalpha_s=\signn(vA^T)$.
By definition of $\Sop$, this means that $\signsalpha_s\in \Sop$.
\epr

We can simplify the definition of $\So$ a bit further, by noticing that the
finite subset $\SV$ can be in fact be generated using only \emph{integer}
vectors.  The definition in~\rref{def:signV}) says that:
\[
\SV = \left\{\sign(\nu \Gamma ) \st \nu  \in  \R^{1\times \ns} \right\} 
\; \subseteq \; \{-1,0,1\}^{1\times \ns}\,.
\]

\bl{lemma:integerS}
\[
\SV = \left\{\sign(\nu \Gamma ) \st \nu  \in  \Z^{1\times \ns} \right\} 
\; \subseteq \; \{-1,0,1\}^{1\times \ns}\,.
\]
\els

\bpr
Pick any $s\in \SV$.  Thus $s=\sign v$, where $v=\nu \Gamma $ for some
$\nu \in \R^{1\times \ns}$.
Consider the set of indices of the coordinates of $v$ that vanish
(equivalently, $s_i=0$), $I=\{i\in \{1,\ldots \ns\} \st v_i=0\}$.
Suppose that $I = \{i_1,\ldots ,i_p\}$.
Let $e_i$ denote the canonical column vector $(0,\ldots 0,,1,0,\ldots 0)^T$ with a
``$1$'' in the $i$th position and zeroes elsewhere, and introduce the
$\ns\times  p$ matrix $E_I = (e_{i_1},e_{i_2},\ldots ,e_{i_p})$.
The definition of $I$ means that $\nu \Gamma E_I = vE_I=0$ and 
$\nu \Gamma e_j=ve_j=v_j\not= 0$ for all $j\not\in I$.
The matrix $D = \Gamma E_I$ has integer, and in particular rational, entries.
Thus, the left nullspace of $D$ has a rational basis, that is, there is a set
of rational vectors $\{u_1,\ldots ,u_q\}$, where $q$ is the dimension of this
nullspace, such that $u_iD=0$ and $uD=0$ if and only if $u$ is a linear
combination of the $u_i$'s.  In particular, since $\nu D=0$, there are real
numbers $r_1,\ldots ,r_q$ such that $\nu =\sum_i r_iu_i$.  Now pick sequences of
rational numbers $r_i^{(k)} \rightarrow  r_i$ as $k\rightarrow \infty $ and define
$\nu ^{(k)}:=\sum_i r_i^{(k)}u_i$.  This sequence converges to $\nu $, and, being
combinations of the $u_i$'s, $\nu ^{(k)}D=0$ for all $k$.
Let $v^{(k)} := \nu ^{(k)}\Gamma $, so we have that $v^{(k)}\rightarrow v$  as $k\rightarrow \infty $,
and $v^{(k)}E_I=0$ for all $k$.
On the other hand, for each $j\not\in I$, as $v e_j\not= 0$, for all large enough $k$,
$(v^{(k)})_j$, the $j$th coordinate of $v^{(k)}$, has the same sign as $v_j$.
In conclusion, for large enough $k$, $\sign v^{(k)} = \sign v = s$.
Multiplying the rational vector $\nu ^{(k)}$ by the least denominator of its
coordinates, the sign does not change, but now we have an integer vector with
the same sign. 
\epr

\section{Summary and implementations}

Our procedure for finding signs $\pil$ of derivatives $\sxl$ consists of the
following steps: 
\ben
\item
Construct a subset ${\cal S}\subseteq \SV$.
\item
For each element $s\in {\cal S}$, test the property $(s^+ \alpha _j)\cdot (s^- \alpha _j) = 0$,
which defines $\SG$.
The $s$'s that pass this test are collected into a set ${\cal S}_G$, which is known
to be a subset of $\SG$.
\item
Take the set of elements of the form 
$\signsalpha_s = \signn(s \alpha )$, for $s$ in ${\cal S}_G$, and add to these
the signs of the rows of the Jacobian $g'$ of $g$
(by assumption, these sign vectors are independent of $x$).
Let us call this set ${\cal T}$.
\item
Now apply the sieve procedure, testing~\rref{eq:conjunction} over elements of
${\cal T}$ (which is a subset of $\So$).
The elements $\pi $ that pass this test are reported as possible signs of
derivatives of steady states with respect to the parameter $\lambda $, in
the sense that they have not been eliminated when
checking~\rref{eq:conjunction} over elements of ${\cal T}$.
\item
If a unique (after eliminating $0$ as well as one element of each pair
$\{\pi ,-\pi \}$) solution remains, we stop.  If there is more than one sign that
passed all tests, and if ${\cal S}$ was a proper subset of $\SV$, we generate a
larger set ${\cal S}$, and hence a potentially larger ${\cal T}$, and repeat the
subsequent steps for the larger subset.
\item
If multiple solutions exist, we may also add additional linear combinations of
those coordinates of $g$ that are linear functions, and enlarge $g$ in that
manner.  (Without loss of generality, arguing in the same manner as for 
$\SV$, we only need to add integer combinations.)
\een

The first step, constructing $\SV$, or a large subset ${\cal S}$ of it, can be done in
various ways.  Since, by Lemma~\ref{lemma:integerS}, we can generate $\SV$
using integer vectors, the elements of $\SV$ have the form $\sign v$
where we may assume, without loss of generality, that each entry of $v=\nu \Gamma $
is either zero or, if nonzero, is either $\geq 1$ or $\leq  -1$.
Thus, testing whether a sign vector $s$ belongs to $\SV$ amounts to testing
the feasibility of a linear program (LP): we need that 
$\nu \Gamma e_i=0$ for those indices $i$ for which $s_i=0$, 
that $\nu \Gamma e_i\leq  -1$  for those indices $i$ for which $s_i=-1$, 
and that $\nu \Gamma e_i\geq 1$  for those indices $i$ for which $s_i=1$.
(These are closed, not strict, conditions, as needed for an LP formulation.)
This means that one can check each of the $3^n$ possible sign vectors
efficiently.

One can combine the testing of LP feasibility with the search over the $3^n$
possible sign vectors into a Mixed Integer Linear Programming (MILP)
formulation, by means of the technique called in the MILP field a ``big M''
approximation.
This is a routine reduction: one first fixes a large positive number $M$, and
then formulates the following inequalities:
\[
\nu \Gamma e_i - M L_i + U_i \leq  0,\quad
-\nu \Gamma e_i - M U_i + L_i \leq  0,\quad
L_i + U_i \leq  1,
\]
where the vector $\nu $ is required to be real and the variables $L_i$, $U_i$
binary ($\{0,1\}$).
Given any solution, we have that
$-M \leq  \nu \Gamma e_i\leq  -1$ (so $s=-1$) for those $i$ for which $(L_i,U_i)=(0,1)$,
$1 \leq  \nu \Gamma e_i\leq  M$ (so $s=1$) for indices for which $(L_i,U_i)=(1,0)$,
and $\nu \Gamma e_i=0$ (i.e., $s_i=0$) when $(L_i,U_i)=(0,0)$.
(This trick will miss any solutions for which $\nu \Gamma e_i\leq  -1$ but
$M$ was not taken large enough that $-M \leq  \nu \Gamma e_i$, or $\nu \Gamma e_i\geq 1$ but
$M$ was not taken large enough that $\nu \Gamma e_i\leq M$.)
The resulting MILP can be solved using relaxation-based cutting plane methods,
branch and bound approaches, or heuristics such as simulated annealing.

Often, however, simply testing sparse integer vectors in the
integer-generating form in Lemma~\ref{lemma:integerS} works well.
In practice, we find that starting with $\nu =\pm e_i^T$ (canonical basis
vectors and their negatives) and sums of pairs of such vectors, in addition to
using the appropriate conservation laws, is typically enough to uniquely
determine the sign vector $\pi $ (up to all signs being reversed, and except for
the trivial solution $\pi =0$), provided that steady states are
uniquely determined from conservation laws.

\section{Example}

\bex{example:ambiguity}
We consider the following reaction network:
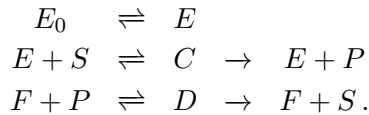
\[
\begin{array}{ccccc}
E_0 & \arrowschem{}{}  & E&\\
E+S & \arrowschem{}{} & C & \arrowchem{}  & E+P\\
F+P & \arrowschem{}{}   & D & \arrowchem{} & F+S\,.
\end{array}
\]
Here $E$ is a kinase that is constitutively activated and inactivated.
Its active form drives a phosphorylation reaction in which a
substrate, $S$ is converted to an active form $P$, which can be
dephosphorylated back into inactive form by a constitutively active
phosphatase $F$.  There are two intermediate enzyme-substrate
complexes as well.
Consider the following three conservation laws:
\be{eq:ss1}
e_0+e+c=e_T
\ee
\be{eq:ss2}
f+d=f_T
\ee
and
\be{eq:ss3}
s+c+p+d = s_T\,.
\ee
We may think of $e_T$ as total amount of enzyme, $f_T$ as total amount of
phosphatase, and $s_T$ as total amount of substrate.
We will study what happens when each of these total amounts is varied while
keeping the other two fixed.
We are also interested in the total concentration of active kinase, free or
bound, $x = e+c$ and the total concentration of product, free or bound,
$y = p+d$.  In order to obtain this information, we add these variables and
add ``virtual'' stoichiometric constraints $p+d-y=0$ and $e_0+x=e_T$
(from~\rref{eq:ss1}) to constrain these variables.

The program returns this outputs:
\begin{verbatim}
    -1    -1     1    -1    -1     1    -1    -1    -1
    e0     e     s     c     d     f     p     x     y
\end{verbatim}
when perturbing only $e_T$,
\begin{verbatim}
    -1    -1     1     1     1     1    -1     1    -1
    e0     e     s     c     d     f     p     x     y
\end{verbatim}
when perturbing only $f_T$, and
\begin{verbatim}
    -1    -1     1     1     1    -1     1     1     1
    e0     e     s     c     d     f     p     x     y
\end{verbatim}
when perturbing only $s_T$.
\eex

\end{document}